\documentclass[conference]{IEEEtran}
\usepackage{graphicx}
\usepackage{epstopdf} 
\usepackage{verbatim} 
\usepackage{multirow} 
\usepackage{multicol}
\usepackage{amsfonts}
\usepackage{bbm}
\usepackage{subfigure}
\usepackage{colortbl} 
\usepackage{indentfirst}
\usepackage{dcolumn,booktabs}
\usepackage{algorithm}
\usepackage[noend]{algpseudocode}
\usepackage{algorithmicx,algorithm}

\usepackage{algpseudocode}
\usepackage{graphics}
\usepackage{epsfig,color}
\usepackage{amsmath}
\usepackage{amssymb}
\ifCLASSINFOpdf

\else
\fi
\newtheorem{Theorem}{Theorem}
\newtheorem{Corollary}{Corollary}

\newtheorem{Lemma}{Lemma}

\begin{document}

\title{Analysis and Optimization of Hybrid Caching in mmWave Networks with BS Cooperation}

\author{
\IEEEauthorblockN{Le Yang, Fu-Chun Zheng and Shi Jin}
}
\maketitle

\begin{abstract}
In this paper, we investigate a hybrid caching strategy maximizing the success transmission probability (STP) in a millimeter wave (mmWave) cache-enabled network. First, we derive theoretical expressions of the STP and the average system transmission delay by utilizing stochastic geometry, then we consider the maximization of the STP and the minimization of the average system transmission delay by optimizing the design parameters. Considering the optimality structure of the NP-hard problem, the original problem is transferred into a multi-choice knapsack problem (MCKP). Finally, we investigate the impact of key network parameters on the STP and the average system transmission delay. Numerical results demonstrate the superiority of the proposed caching strategy over the conventional caching strategies in the mmWave cache-enabled networks.
\end{abstract}

\begin{IEEEkeywords}
Caching strategy, stochastic geometry, millimeter wave, delay, successive interference cancellation.
\end{IEEEkeywords}

%
\IEEEpeerreviewmaketitle

\section{Introduction}
Due to the rapid proliferation of various multi-media applications and smart mobile devices, the mobile data traffic has witnessed an unprecedented growth and imposed heavy burden on the backhaul links. Caching the popular contents at the BSs has become a promising way to alleviate the burden of the backhaul link. The utilization of the millimeter wave (mmWave) band is another technique to meet the ever-increasing demand of the data traffic for future networks due to that the available spectrum at these frequencies can be 200 times greater than all cellular allocations today that are largely constrained to the sub-6GHz prime RF real estate \cite{millimeter-for-5G}.

Compared with the sub-6GHz networks, the mmWave networks have two fundamental differences: the sensitivity to blockages and the propagation loss \cite{channel-modeling}. Fortunately, the beamforming gain of directional antenna arrays can be utilized to compensate the propagation loss experienced by the receiver, resulting in comparable coverage ranges \cite{Bai}. In fact, several channel measurements have revealed that the transmission range of 150-200m can be obtained in mmWave bands \cite{measurement-1}\cite{measurement-2}.

The caching strategy design in the cellular networks has been studied in the existing works utilizing stochastic geometry as the analyzing tool. Note that caching strategy design is a prerequisite for the file dissemination due to the limited cache size. In addition, the file diversity affects the performance of a caching strategy. In \cite{origin}, the outage probability and the average delivery rate of the cache-enabled networks were analyzed. In \cite{cui}--\cite{tier-level}, the analytical expression of the successful transmission probability (STP) for the sub-6GHz cache-enabled heterogeneous networks was derived. In particular, the authors in \cite{N-tier-caching} obtained the optimal caching probability (maximizing the STP in the interference-limited regime) as well as investigating the tradeoff between the BS density and the cache size under the uniform caching strategy. In \cite{tier-level}, the caching probability optimization problem in the sub-6GHz multi-tier cache-enabled networks with a non-uniform SIR threshold was proved to be non-convex and the sub-optimal caching probabilities were obtained. Furthermore, the caching probability optimization problem with a uniform SIR threshold was convex and the corresponding closed-form solution was achieved. In \cite{caching-towards-maximum}, the authors studied the optimal caching policy which respectively maximizes the STP and the area spectral efficiency (ASE) in the sub-6GHz two-tier cache-enabled networks. In \cite{wen}, the expressions of STP under two cooperative transmission strategies in the sub-6GHz cache-enabled networks were derived. In addition, locally optimal caching probabilities were achieved in the general case and the globally optimal caching probability were provided in the low data rate case.

The analysis was also extended to the mmWave or hybrid networks. In \cite{UK-sub-6}, a cross-entropy optimization method was proposed to obtain the sub-optimal performance in a hybrid cache-enabled network. In addition, a heuristic file placement scheme was provided to achieve the balance between the transmission reliability and the content diversity. In \cite{UK-mmWave}, the performance of a hybrid small-cell network where the SBSs was overlaid by the backhaul-connected SBSs was analyzed and the most popular caching (MPC) strategy was proved to be the optimal caching strategy for achieving the maximal ASE for the case of high Zipf skewedness factors. The performance of the hybrid cache-enabled heterogeneous networks utilizing the LOS ball model was analyzed and the impact of critical physical layer parameters on the network performance was examined in \cite{Yi-cache-mmWave}. The authors \cite{origin}-\cite{Yi-cache-mmWave} considered the random caching strategy where the files were wholly cached in the BSs. The main drawback of the random caching strategy lies in that the serving distance becomes larger when the file diversity increases. Therefore, the advantage of the file diversity cannot be fully utilized.

When a user is served by multiple BSs, a file may be divided into multiple subfiles and each subfile can be cached in a BS. Hence, the distance between the user and its serving BS may be reduced, leading to the improvement of the file diversity. In \cite{network-coding}, the network coding-based caching strategy was proposed and the cache miss probability was analyzed. In \cite{MDS}, the maximum distance separable (MDS) code-based caching strategy was proposed and the total number of packet transmissions was minimized. In \cite{partition-based}, a partition-based caching strategy was proposed and the successful content delivery probability was analyzed. Note that successive interference cancelation (SIC) was adopted at each user to decode the subfiles of its requested file. With the application of the SIC, the exploitation of the file diversity was facilitated and the performance of the cache-enabled networks was improved. The utilization of SIC depends on the imbalance of received signal powers from the transmitters at different locations. The idea of SIC is to decode the signals according to the decreasing signal power and subtract the decoded signal from the received signals. The process continues until all the signals are decoded. The benefit of SIC in the large-scale wireless networks was revealed by utilizing tools from stochastic geometry. In \cite{power}, the SIC based on power order was considered and the bounds on the successful decoding probability was derived. In \cite{distance-1} and \cite{distance-2}, the SIC based on the distance order was considered and the the closed-form expressions for the coverage probabilities were obtained in D2D and heterogeneous networks, respectively.

While the partition-based caching strategies in the sub-6GHz networks has been investigated, the partition-based strategy in the mmWave networks still remains to be studied. Furthermore, the advantage of the BS cooperation has not been utilized in the existing works. Note that the benefit of BS cooperation in the mmWave networks was demonstrated in \cite{5G-CoMP} and \cite{BS-cooperation-mmwave}. Specifically, the performance of the coordination multipoint (CoMP) and macro-diversity was analyzed. The measurement quantitatively showed that, compare to a user served by a single BS, the outage probability of the user served by multiple BSs was significantly reduced. Therefore, we consider a hybrid caching strategy in the mmWave networks. The files with higher popularity are cached wholly and the joint transmission (JT) strategy is utilized to provide transmission reliability. The files with lower popularity are partitioned and coded cached and the parallel transmission (PT) strategy is utilized to provide file diversity. By utilizing the hybrid caching strategy, we can strike a balance between the transmission reliability and file diversity. The main contribution of this paper are summarized as follows:
\begin{enumerate}
\item By utilizing the LOS probability function to model the LOS/NLOS state of each link, we characterize the distribution of the path loss between $u_0$ and its serving BS. We derive the expression of the STP in the mmWave cache-enabled networks under the JT and PT strategies. In addition, a simplified expression of the STP is obtained under a special case where the blockage parameter is sufficiently small. Moreover, by taking the backhaul delay into consideration, the expressions of the average system transmission delays for the JT, PT and uncached transmission (UT) strategies in the mmWave cache-enabled networks are obtained.

\item We consider the maximization of the STP or the minimization of the average system transmission delay by optimizing the design parameters. By exploiting the optimality structure of the NP-hard problem, the optimization problem is transformed into a multi-choice knapsack problem (MCKP) and a near optimal solution can be obtained with 1/2 approximation guarantee. In addition, the we consider the special case where the blockage parameter and the file size are both sufficiently small and obtained a closed-form asymptotically optimal solution.

\item We reveal the effect of the physical layer parameters and the design parameters on the STP and the average system transmission delay. Numerical results demonstrates the superiority of the proposed method over the existing caching strategies and shows that the advantage of the cooperative case over the non-cooperative case is reduced when the blockage parameter increases.
\end{enumerate}

The reminder of this paper is organized as follows. In Section $\text{\uppercase\expandafter{\romannumeral2}}$, the system model is introduced. In Section $\text{\uppercase\expandafter{\romannumeral3}}$, the STP and local delay of the mmWave cache-enabled network is derived. In Section $\text{\uppercase\expandafter{\romannumeral4}}$, the maximization of the STP and the minimization of the average system transmission delay are obtained by optimizing the design parameters. In Section $\text{\uppercase\expandafter{\romannumeral5}}$, numerical results are presented to demonstrate the superiority of the proposed algorithm over the existing schemes. In addition, the effect of the key network parameters on the STP and the average system transmission delay are presented. Finally, conclusions are provided in Section $\text{\uppercase\expandafter{\romannumeral6}}$.

\section{System model}
\begin{figure}
  \centering
  \includegraphics[width=3.5in]{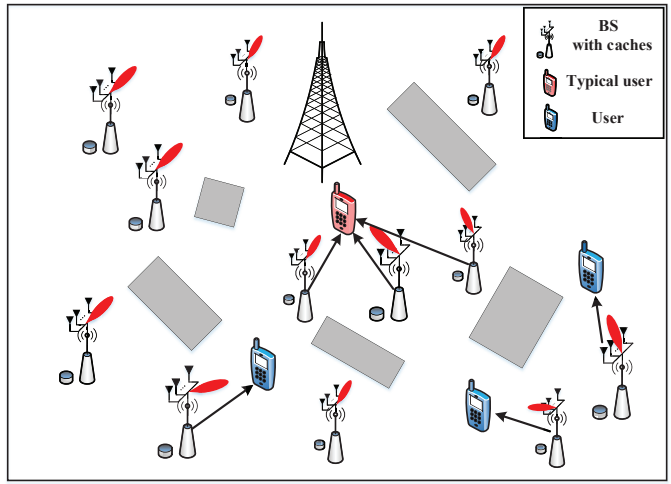}
  \caption{The layouts of a mmWave cache-enabled network.}\label{framework}
\end{figure}

We consider a downlink cache-enabled network, as shown in Fig. \ref{framework}. The locations of the BSs and the gateways are assumed to follow the independent homogeneous Poisson point processes (PPP) $\Phi$ with density $\lambda$ and $\Phi_{\text{G}}$ with $\lambda_{\text{G}}$, respectively. Let $P$ denote the transmit power of each BS. We assume that the BSs operate over the mmWave frequency band. The available bandwidth is denoted by $W$ (in Hz). We consider a discrete time system where the time is slotted with equal duration, i.e., $T$ seconds. Without loss of generality, according to Slivnyak's theorem \cite{Bai}, we can study the performance of a typical user $u_0$ located at the origin $o\in\mathbb{R}$.

Both large-scale and small-scale fading are considered. Depending on the visibility of the BS to $u_0$, the BS can be either line-of-sight (LOS) or non-line-of-sight (NLOS). That is, if there are no blockages between $u_0$ and the BS, the corresponding link is considered to be LOS. Otherwise, the link is NLOS. A LOS probability function, i.e., $p(x)=e^{-\beta x}$, is utilized to model the LOS probability of an arbitrary link between the BS at the distance $x$ and $u_0$. Note that the blockage parameter $\beta$ is dependent on the blockages and BS density \cite{Bai}. Accordingly, the NLOS probability of the corresponding link is therefore $1-e^{-\beta x}$. The desired signal received by $u_0$ is attenuated by the path loss. To distinguish the LOS/NLOS states of the links, different path loss laws are applied to the links under the LOS/NLOS link states, which are given by
\begin{equation}
L(x)=
\begin{cases}
\kappa_\text{LOS}x^{\alpha_{\text{LOS}}}, \quad\text{with prob.}\quad e^{-\beta x}\\
\kappa_\text{NLOS}x^{\alpha_{\text{NLOS}}}, \quad\text{with prob.}\quad 1-e^{-\beta x}
\end{cases}
\end{equation}
where $\kappa_{\text{LOS}}$ and $\kappa_{\text{NLOS}}$ are the path losses of the LOS/NLOS links at the reference distance of 1 meter, $\alpha_{\text{LOS}}$ and $\alpha_{\text{NLOS}}$ represent the path loss exponents of the LOS and NLOS links, respectively.

The small-scale fading is assumed to be independent Nakagami fading for all BSs and the Nakagami parameters is denoted by $M$. Therefore, $|h|^2$ is a normalized Gamma variable. Note that when $\nu_{\text{LOS}}=\nu_{\text{NLOS}}=1$, the small-scale fading reduces to Rayleigh distribution.

Let $\mathcal{F}\triangleq \{1,2,\cdots,F\}$ denote a set of $F$ files in the core network. Assume that all the files has the same size of $S$ bits. \footnote{Note that the results can easily be extended to the case where the contents have different file sizes (e.g. by combining multiple files of different sizes to form files of equal size or splitting files of different sizes into segments of equal size) \cite{cui}.}, and the file popularity distribution is identical among all users. Time is divided into slots with equal duration $T$. Let $R=S/T$ denote the target data rate for a successful file delivery. Let $p_f$ denote the probability that File $f$ is requested by a user, i.e., the popularity of File $f$ is $p_f$, where $\sum_{f=1}^{F}p_f=1$. In addition, we can always assume that $p_1\ge p \ge \cdots \ge p_F$. Hence, the file popularity distribution can be expressed as $\mathbf{p}\triangleq \{p_1,p,\cdots,p_F\}$, which is assumed to be known a priori. Note that the popularity of different files evolves at a relatively slow timescale and can be estimated in practice (e.g. by the machine learning \cite{machine-learning}). Each BS is equipped with a cache of size $C \le F$ to store $C$ different files from $\mathcal{F}$.

\subsection{Performance Metric}
Two phases, i.e., file placement phase and the file delivery phase, are considered. In the file placement phase, the files are retrieved from the core network and placed into the BS by utilizing the hybrid strategy in the off-peak time. For the MPC strategy, the files are wholly cached in the BSs. For the LDC strategy, the files are partitioned into subfiles and coded cached in the BSs. Compared with the MPC strategy, the LDC strategy provides a larger file diversity since more files are cached in the BSs. However, the transmission reliability for the LDC strategy degrades since different subfiles of the requested file needs to be decoded.

Assume that $N$ is the SIC capability of $u_0$. That is, the decoding and cancellation can be performed at most $N$ times to obtain the desired signal. Let $\bar{\mathcal{N}}\triangleq\{2,3,\cdots,N\}$, $\mathcal{N}=\bar{\mathcal{N}}\cup\{0,1\}$. In addition, denote by $s_f\in\mathcal{N}$ the caching status of File $f$ and we have
\begin{equation}\label{constraint-1}
s_f\in\mathcal{N},f\in\mathcal{F}.
\end{equation}

Note that all the subfiles are coded by utilizing random linear network coding (RLNC) \cite{RLNC}. If $s_f=1$, File $f$ will wholly cached in $N$ cooperative BSs. If $s_f=0$, File $f$ will be not cached in the BSs and should be retrieved from the core network through the backhaul link. If $s_f\in\bar{\mathcal{N}}$, File $f$ will be partitioned into subfiles and coded cached in the BSs. Note that the BS cache size satisfies that
\begin{equation}\label{constraint-2}
\sum_{f=1}^{F}\frac{1}{s_f}\leq C.
\end{equation}

In the file delivery phase, File $f$ is transmitted to $u_0$ from the cooperative BSs. If the transmission fails, File $f$ will be retransmitted. Note that the retransmission continues until a successful transmission occurs. Different transmission strategies are applied according to the caching status, which is shown as follows:

\emph{1) Joint Transmission:} If File $f$ with $s_f=1$ is requested, it will be simultaneously transmitted from $N$ BSs to enhance the transmission reliability.

\emph{2) Parallel Transmission:} If File $f$ with $s_f\in\bar{\mathcal{N}}$ is requested, different subfiles of File $f$ will be transmitted from the cooperative $s_f$ BSs simultaneously. SIC is utilized to increase the SINR at $u_0$, where the subfiles are decoded according to the distance and the subfiles from nearer BSs are decoded first and then removed from the received subfiles at $u_0$.

\emph{3) Uncached Files Transmission:} If File $f$ with $s_f=0$ is requested, File $f$ will be retrieved by the BS providing the largest signal power from the core network through the backhaul link.

Note that, in this paper, we assume that the antenna arrays at the BSs perform directional beamforming, where the main lobes are directed towards the dominant propagation paths while smaller side lobes direct power into other directions. In addition, for analytical tractability, as in \cite{Bai}, array patterns are approximated by the sectored antenna model where the beam direction of the interfering links is assumed to follow a uniform distribution within $\left[0,2\pi\right]$. The effective antenna gain between a BS in tier $k$ and $u_0$ is a discrete random variable described by
\begin{equation}
G=
\begin{cases}
M_T, &\text{with probability}\quad \frac{\theta_T}{2\pi},\\
m_T, &\text{with probability}\quad \frac{2\pi-\theta_T}{2\pi},
\end{cases}
\end{equation}
where $\theta_T$ denote the half power beamwidth of the antenna arrays deployed at the BS and the users, $M_T$, $m_T$ the main lobe gain and side lobe gain of the antenna arrays deployed the BSs. Assume that the perfect CSI is available at the BS and the antenna array is steerable to guarantee that the maximum array gain can be exploited with the aid of perfect estimation of channels (the channel between $u_0$ and its serving BS lies in the boresight direction of the antennas of both BSs and $u_0$).

Next, we analyze the expression of SINR under two transmission strategies. Under the JT strategy, the instantaneous signal-to-interference-and-noise ratio (SINR) is given by (\ref{SINR-pico}), as shown on the top of the next page.
\begin{equation}\label{SINR-pico}
\text{SINR}^{\text{JT}}
=\frac{\sum_{x_i\in\mathcal{C}}PG_{x_i}|h_{x_i}|^2L^{-1}(x_i)}{\sum_{x_i \in\Phi \setminus\mathcal{C}} P G_{x_i} |h_{x_i}|^2 L^{-1}(x_i)+\sigma^2},
\end{equation}
where $G_{x_i},x_i\in\mathcal{C}$ denotes the antenna gain of the cooperative serving links between the BS at $x_i$ and $u_0$, $n_{0}\sim\mathcal{CN}(0,\sigma^2)$ the complex additive white Gaussian noise (AWGN).

Under the PT strategy, the instantaneous SINR for the $n$th transmission can be expressed as:
\begin{equation}\label{SINR-pico}
\text{SINR}_n^{\text{PT}}
=\frac{PG_{x_n}|h_{x_n}|^2L^{-1}(x_{n})}{\sum_{x_i \in\Phi \setminus\mathcal{B}(0,x_n)} P G_{x_i} |h_{x_i}|^2 L^{-1}(x_i)+\sigma^2}.
\end{equation}

In this paper, we adopt the STP \cite{cui} and the average system transmission delay \cite{system-transmission-delay} as the performance metrics. First, we provide the definition of the STP. If $s_f=1$, the JT strategy is employed. Under the JT strategy, the transmission of File $f$ is considered to be successful if the data rate exceeds a pre-defined threshold $T$. If $s_f\in\bar{\mathcal{N}}$, the PT strategy is employed. Under the PT strategy, the transmission of File $f$ is considered to be successful only if all the $s_f$ transmissions succeed. Based on the above analysis, the STP of the cache-enabled mmWave networks is given by
\begin{equation}
\mathcal{P}_f=\\
\begin{cases}
\mathbbm{P}\left[W\left(1+\text{SINR}^{\text{JT}}\right)>\frac{S}{s_fT}\right],&\text{if}\ s_f=1,\\
\mathbbm{P}\left[\bigcap\limits_{n=1,\cdots,s_f}W\left(1+\text{SINR}^{\text{PT}}_n\right)>\frac{S}{s_fT}\right],&\text{if}\ s_f\in\bar{\mathcal{N}}.
\end{cases}
\end{equation}

Next, we provide the definition of the average system transmission delay. If $s_f\in\mathcal{N}\backslash\{0\}$, the average system transmission delay equals the mean local delay. We denote the mean local delay by $D$. Conditioned on $\Phi$, the local delay is geometrically distributed with parameter $\mathcal{P}(\theta|\Phi)$, we then have
\begin{equation}
\mathbbm{P}[D=d\mid\Phi]=(1-\mathcal{P}(\theta\mid\Phi))^{d-1}\mathcal{P}(\theta\mid\Phi),
\end{equation}
for $d=1,\cdots$. The mean of the geometrically distributed random variable $D$ conditioned on $\Phi$ is $\mathbbm{E}[D\mid\Phi]=\frac{1}{\mathcal{P}(\theta\mid\Phi)}$. The mean local delay can then be obtained by calculating the expectation with respect to $\Phi$ as
\begin{equation}\label{delay-definition}
\mathbbm{E}[D_f]=\mathbbm{E}_{\Phi}\left[\mathbbm{E}\left[D_f\mid\Phi\right]\right]=\mathbbm{E}_{\Phi}\left[\frac{1}{\mathcal{P}_f(\theta\mid\Phi)}\right].
\end{equation}

If $s_f=0$, the average system transmission delay equals the summation of the backhaul delay and the transmission delay from the BS providing the strongest signal power to $u_0$. Then the average system transmission delay is
\begin{equation}\label{delay-definition}
\mathbbm{E}[D_f]=
\begin{cases}
\mathbbm{E}_{\Phi}\left[\left.\mathbbm{P}\left[W\left(1+\text{SINR}^{\text{JT}}\right)>\frac{S}{s_fT}\right|\Phi\right]^{-1}\right],\\ \qquad\qquad\qquad\qquad\qquad\qquad\text{if}\ s_f=1,\\
\mathbbm{E}_{\Phi}\left[\mathbbm{P}\left[\left.\bigcap\limits_{n=1,\cdots,s_f}W\left(1+\text{SINR}^{\text{PT}}_n\right)>\frac{S}{s_fT}\right|\Phi\right]^{-1}\right],\\
\qquad\qquad\qquad\qquad\qquad\qquad\text{if}\ s_f\in\bar{\mathcal{N}},\\
D^{\text{UT}},\qquad\qquad\qquad\qquad\quad\quad\text{if}\ s_f=N+1,
\end{cases}
\end{equation}
where $D^{\text{UT}}$ denotes the average system transmission delay for the files not cached in the BSs.

In this paper, our goal is therefore to obtain the maximization of the conditional STP or the minimization of the average system transmission delay by optimizing the design parameters.

\section{Analysis of STP and local delay}
In this section, we provide the STP and delay of the cache-enabled networks. Since the files may be non-partitioned cached, coded cached or uncached. The expressions of STP and delay can be derived based on different caching status.

\subsection{Analysis of STP}
In this subsection, we provide the expressions of STP under the JT and PT strategy, respectively.

\subsubsection{JT Strategy}
When File $f$ is non-partitioned and cached in $N$ BSs, it will be jointly transmitted from $N$ BSs. Before we provide the expression of the STP under PT strategy, we first derive the joint probability density function (PDF) of the path loss between $u_0$ and its serving BSs in the following lemma.
\begin{Lemma}
For the JT strategy, the joint PDF of the path loss between $u_0$ and its serving BSs is given by
\begin{equation}
f_{L}(x)=\prod_{n=1}^{N}\Lambda^{'}([0,x_n))\exp(-\Lambda([0,x_N))),
\end{equation}
where
\begin{equation}\label{Lambda}
\begin{split}
&\Lambda([0,x))=\pi\lambda(x/\kappa_{\text{NLOS}})^{2/\alpha_{\text{NLOS}}}-2\pi\lambda\beta^{-2}\\
&\left(1-e^{-\beta(x/\kappa_{\text{NLOS}})^{1/\alpha_{\text{NLOS}}}}
\left(1+\beta(x/\kappa_{\text{NLOS}})^{1/\alpha_{\text{NLOS}}}\right)\right)\\
&+2\pi\lambda\beta^{-2}\left(1-e^{-\beta(x/\kappa_{\text{LOS}})^{1/\alpha_{\text{LOS}}}}
\left(1+\beta(x/\kappa_{\text{LOS}})^{1/\alpha_{\text{LOS}}}\right)\right),
\end{split}
\end{equation}
\begin{equation}\label{Lambda-derivative}
\begin{split}
&\Lambda^{'}([0,x))
=\frac{2\pi\lambda }{\alpha_{\text{LOS}}\kappa_{\text{LOS}}^{\frac{2}{\alpha_{\text{LOS}}}}}x^{\frac{2}{\alpha_{\text{LOS}}}-1}
e^{-\beta (\frac{x}{\kappa_{\text{LOS}}})^{\frac{1}{\alpha_{\text{LOS}}}}}\\
&+\frac{2\pi\lambda }{\alpha_{\text{NLOS}}\kappa_{\text{NLOS}}^{\frac{2}{\alpha_{\text{NLOS}}}}}x^{\frac{2}{\alpha_{\text{NLOS}}}-1}
\left(1-e^{-\beta (\frac{x}{\kappa_{\text{NLOS}}})^{\frac{1}{\alpha_{\text{NLOS}}}}}\right).
\end{split}
\end{equation}
\end{Lemma}

\emph{Proof:} See Appendix A.

Next, the expression of the STP under the JT strategy is provided in the following theorem.
\begin{Theorem}\label{theorem-STP}
The STP of the cache-enabled mmWave networks under the JT strategy is given by
\begin{equation}\label{STP-pico}
\begin{split}
\mathcal{P}_f=\idotsint\limits_{0<l_{1}<\cdots<l_{N}<\infty}\lVert\exp\{\Omega_{MN}\}\rVert_1f_{L}(l)\text{d}l_{1}\cdots\text{d}l_{N},
\end{split}
\end{equation}
where $\exp(\mathbf{A})$ is the matrix exponential and $\exp(\mathbf{A})=\sum_{k=0}^{\infty}\frac{\mathbf{A}^{k}}{k!}$. $\lVert\cdot \rVert$ denotes the $l_1$-induced norm. $N_{MN}$ is the toeplitz matrix, which can be expressed as
\begin{equation}
\Omega_{MN}=
\begin{bmatrix}
\omega_{0}\\
\omega_1&\omega_0\\
\vdots&\vdots&\ddots\\
\omega_{MN-1}&\omega_{MN-2}&\cdots&\omega_0
\end{bmatrix},
\end{equation}
where $\omega_0$ and $\omega_j$ are given by (\ref{omega-0}) and (\ref{omega-j}), as shown on the top of the next page.

\begin{equation}\label{omega-0}
\begin{split}
\omega_0=-\sum_G\left(\int_{l_N}^{\infty}
\left(1-\left(1+\frac{G}{xG_0M}\right)^{-M}\right)\Lambda(\text{d}x)\right),
\end{split}
\end{equation}
\begin{equation}\label{omega-j}
\begin{split}
\omega_j&=\sum_{G}\left(\frac{G(M)_j\left(2^{\frac{S}{s_fT}}-1\right)}{j!MG_0\sum_{n=1}^{N}l_n^{-1}}\int_{0}^{\frac{TG}{MG_0l_N\sum_{n=1}^{N}l_n^{-1}}}
\frac{t^{j-2}}{(1+t)^{M+j}}\right.\\
&\left.\Lambda^{'}\left(0,\frac{G\left(2^{\frac{S}{s_fT}}-1\right)}{MG_0\sum_{n=1}^{N}l_n^{-1}}\right)\text{d}t\right)-\frac{\sigma^2}{PG_0}\max\{2-j,0\}.
\end{split}
\end{equation}
\end{Theorem}
\emph{Proof:} See Appendix B.

It can be observed from Theorem \ref{theorem-STP} that the expression of STP is in a complex form including multiple integrals. The main challenge in analyzing the mmWave networks is to tackling the high order derivatives of the Laplace transforms of the noise and interference. The Fa\`a di Bruno formula is employed in \cite{Faa-di-Bruno} in order to evaluate the performance of the mmWave networks. However, the complexity of the analytical expression is high since a large number of products and summations are included. Therefore, we utilize the method in \cite{recursive-1}, \cite{recursive-2} with which the STP is expressed by the $l_1$-induced norm of a Toeplitz matrix.

The STP is dependent on two sets of network parameters: the physical layer parameters (i.e., the BS density $\lambda$ and the transmit power $P$) and the content-related parameters, (i.e., the caching status $\mathbf{s}\triangleq[s_1,\dots,s_F]^{\text{T}}$). Note that the STP under the JT strategy is a monotonically increasing function of the Nakagami parameter and the number of cooperative BSs. For the Nakagami parameter, an intuitive explanation is that the fading is less severe with the increasing Nagakami parameter, which results in a better STP. For the number of cooperative BSs, the reason is that the signal strength can be enhanced when more BSs are included into the transmission process, leading to a better STP.

The expression in Theorem \ref{theorem-STP} requires numerical evaluation of multiple integrals. In order to simplify the expression, we consider a special case where the blockage parameter is sufficiently small and the Nakagami parameter $M=1$.

\begin{Corollary}\label{corollary-STP}
When $\beta\rightarrow 0$ and $M=1$, the STP of the cache-enabled networks under the JT strategy is given by
\begin{equation}
\begin{split}
\mathcal{P}_{f,\infty}^{\text{JT}}=\idotsint\limits_{0<u_{1}<\cdots<u_{N}<\infty}\exp\left(-u_NF(\alpha,T)\right)\text{d}u_{1}\cdots\text{d}u_{N},
\end{split}
\end{equation}
where
\begin{equation}
\begin{split}
&F(\alpha,T)\\
&={}_2F_1\left(-\frac{2}{\alpha_{\text{LOS}}},1;1-\frac{2}{\alpha_{\text{LOS}}};-\frac{2^{\frac{S}{s_fT}}-1}{\sum_{n=1}^{N}\left(\frac{u_N}{u_1}\right)^{\frac{\alpha_{\text{LOS}}}{2}}}\right)-1,
\end{split}
\end{equation}
\end{Corollary}

\emph{Proof:} See Appendix C.

\subsubsection{PT strategy}
When File $f$ is partitioned and coded cached in $s_f$ BSs, $s_f$ subfiles will be transmitted simultaneously by parallel streams. Before we provide the expression of the STP under PT strategy, we first derive the PDF of the path loss between $u_0$ and its $n$th cooperative serving BS in the following lemma. From \cite{distance-distribution}, we can obtain the distribution of the path loss between $u_0$ and its $n$th serving BS in the following lemma.

\begin{Lemma}
For the PT strategy, the PDF of the path loss between $u_0$ and the BS with the $n$th smallest path loss is given by
\begin{equation}
f_{L_n}(x)=\frac{(\Lambda([0,x)))^{n-1}\Lambda^{'}([0,x))}{(n-1)!}\exp(-\Lambda([0,x)))
\end{equation}
where $\Lambda([0,x))$ and $\Lambda^{'}([0,x))$ are given by (\ref{Lambda}) and (\ref{Lambda-derivative}), respectively.
\end{Lemma}

Next, the expression of the STP under the PT strategy is provided in the following theorem.
\begin{Theorem}
The STP of File $f$ in the cache-enabled mmWave networks under the PT strategy is given by
\begin{equation}\label{distance-PT}
\begin{split}
&\mathcal{P}_f^{\text{PT}}=\prod_{n=1}^{s_f}\int_{0}^{\infty}\lVert\exp\{\Omega_{M}\}\rVert_1f_{L_n}(l)\text{d}l,
\end{split}
\end{equation}
where $\exp(\mathbf{A})$ is the matrix exponential and $\exp(\mathbf{A})=\sum_{k=0}^{\infty}\frac{\mathbf{A}^{k}}{k!}$. $\lVert\cdot \rVert$ denotes the $l_1$-induced norm. $N_{M}$ is the toeplitz matrix, which can be expressed as
\begin{equation}
\Omega_{M}=
\begin{bmatrix}
\omega_{0,n}\\
\omega_{1,n}&\omega_{0,n}\\
\vdots&\vdots&\ddots\\
\omega_{M-1,n}&\omega_{M-2,n}&\cdots&\omega_{0,n}
\end{bmatrix},
\end{equation}
where $\omega_{0,n}$ and $\omega_{j,n}$ are given by (\ref{omega-n-0}) and (\ref{omega-n-j}), as shown on the top of the next page.

\begin{equation}\label{omega-n-0}
\begin{split}
\omega_{0,n}=&-\sum_G\int_{l_n}^{\infty}
\left(1-\left(1+\frac{G}{xG_0M}\right)^{-M}\right)\Lambda(\text{d}x),
\end{split}
\end{equation}

\begin{equation}\label{omega-n-j}
\begin{split}
\omega_{j,n}=&\sum_{G}\frac{Gl_n(M)_j\left(2^{\frac{S}{s_fT}}-1\right)}{j!MG_0}\int_{0}^{\frac{TG}{MG_0}}
\frac{t^{j-2}}{(1+t)^{M+j}}\\
&\Lambda^{'}\left(0,\frac{Gl_n\left(2^{\frac{S}{s_fT}}-1\right)}{MG_0}\right)\text{d}t-\frac{\sigma^2}{PG_0}\max\{2-j,0\}.
\end{split}
\end{equation}
\end{Theorem}

\emph{Proof:} See Appendix D.

We consider a special case where the blockage parameter is sufficiently small and the Nakagami parameter $M=1$.

\begin{Corollary}\label{corollary-STP}
When $\beta\rightarrow 0$ and $M=1$, the STP of the cache-enabled networks under the PT strategy is given by
\begin{equation}
\begin{split}
\mathcal{P}_{f,\infty}^{\text{PT}}=\frac{1}{{}_2F_1\left(\frac{2}{\alpha_{\text{LOS}}},1;1-\frac{2}{\alpha_{\text{LOS}}};1-2^{\frac{S}{s_fT}}\right)^{\frac{s_f(s_f+1)}{2}}}.
\end{split}
\end{equation}
\end{Corollary}

\emph{Proof:} When $\beta\rightarrow 0$ and $\nu_{\text{LOS}}=1$, the PDF of $f_{l_n}(x)$ reduces to
\begin{equation}
f_{l_n}(x)=\frac{2\pi^n\lambda^n}{\alpha_{\text{LOS}}\kappa_{\text{LOS}}^{\frac{2n}{\alpha_{\text{LOS}}}}}x^{\frac{2n}{\alpha_{\text{LOS}}}-1}
\exp\left(-\pi\lambda\left(\frac{x}{\kappa_{\text{LOS}}}\right)^{\frac{2}{\alpha_{\text{LOS}}}}\right)
\end{equation}

When the $n$th subfile of File $f$ is transmitted by the BS, the STP $\mathcal{P}_n$ can be derived as
\begin{equation}
\begin{split}
&\mathcal{P}_{f,n,\infty}=
\int_{0}^{\infty}f_{l_n}(x)\pi\lambda \left(\frac{x}{\kappa_{\text{LOS}}}\right)^{\frac{2}{\alpha_{\text{LOS}}}}\\
&\left({}_2F_{1}\left[-\frac{2}{\alpha_{\text{LOS}}},M;1-\frac{2}{\alpha_{\text{LOS}}};\frac{G\left(2^{\frac{S}{s_fT}}-1\right)}{MG_0\sum_{n=1}^{N}l_n^{-1}}\right]-1\right)\text{d}x\\
&=\frac{1}{{}_2F_1\left(\frac{2}{\alpha_{\text{LOS}}},1;1-\frac{2}{\alpha_{\text{LOS}}};1-2^{\frac{S}{s_fT}}\right)^{s_f}}.
\end{split}
\end{equation}

Then the STP of File $f$ the cache-enabled networks under the PT strategy is
\begin{align}
\mathcal{P}_{f,\infty}&=\prod_{n=1}^{s_f}\mathcal{P}_{f,n,\infty}\\
&=\frac{1}{{}_2F_1\left(\frac{2}{\alpha_{\text{LOS}}},1;1-\frac{2}{\alpha_{\text{LOS}}};1-2^{\frac{S}{s_fT}}\right)^{\frac{s_f(s_f+1)}{2}}}.
\end{align}

\subsection{Analysis of Average System Transmission Delay}
The average system transmission delays are different for the files cached in the BSs and those not cached (e.g., \cite{Yaru}). When $u_0$ requests a file which happens to be cached in the BS, $u_0$ can directly obtain the corresponding file from the cooperative serving BSs by utilizing the MPC or LDC strategy in which case only the local delay is involved. However, when the requested file is not cached in any BS, $u_0$ needs to retrieve the corresponding file from the core network through the backhaul. Hence, the average system transmission delay now includes both the local delay and the backhaul delay.

\subsubsection{JT Strategy}
When File $f$ is wholly cached, it will be transmitted jointly from the $N$ cooperative BSs. According to \ref{delay-definition}, the mean local delay under the JT strategy can be obtained in the following theorem.

\begin{Theorem}
The local delay of File $f$ in the cache-enabled mmWave networks under the JT strategy is given by
\begin{equation}
\begin{split}
&D^{\text{JT}}=\idotsint\limits_{0<l_{1}<\cdots<l_{N}<\infty}\lVert\exp\{\Omega_{MN}\}\rVert^{-1}_1f_{L}(l)\text{d}l_{1}\cdots\text{d}l_{N}.
\end{split}
\end{equation}
\end{Theorem}

\subsubsection{PT Strategy}
When File $f$ is partitioned and coded cached in $s_f$ BSs ($s_f\in\bar{\mathcal{N}}$), $s_f$ subfiles will be transmitted simultaneously by parallel streams. According to (\ref{delay-definition}), the average system transmission delay of File $f$ can be obtained in the following theorem.

\begin{Theorem}
The mean local delay of File $f$ in the cache-enabled mmWave networks under the PT strategy is given by
\begin{equation}
\begin{split}
&D_f^{\text{PT}}=\prod_{n=1}^{s_f}\int_{0}^{\infty}\lVert\exp\{\Omega_{M}\}\rVert^{-1}_1f_{L_n}(l)\text{d}l.
\end{split}
\end{equation}
\end{Theorem}

\subsubsection{UT strategy}
When File $f$ is uncached ($s_f=0$), it will be retrieved from the core network through the backhaul link. Let $D_b$ be the backhaul delay, the average system transmission delay is given by
\begin{equation}
D_b=\frac{1}{2}\upsilon\lambda\lambda_{\text{G}}^{-3/2},
\end{equation}
where $\upsilon$ denotes a scaling factor of the backhaul infrastructure. Then we obtain the average system transmission delay as follows:
\begin{equation}\label{UT-delay}
D^{\text{UT}}=D_b+\int_{0}^{\infty}\lVert\exp\{\Omega_{MN}\}\rVert_1f_{L}(l)\text{d}l,
\end{equation}
where the latter term of (\ref{UT-delay}) is obtained by substituting $N=1$ into (\ref{STP-pico}).

\section{Optimization of STP and Average System Transmission Delay}
The caching design affects the STP and the average system transmission delay of the mmWave cache-enabled networks. Therefore, we aim to maximize the STP and the average system transmission delay by optimizing $\mathbf{s}$.

\subsection{Optimization of STP}
In this subsection, we would like to maximize the STP by optimizing $\mathbf{s}$.

\emph{Problem 1 (Optimization of STP):}
\begin{equation}\label{problem-1-equation}
\max\limits_{\mathbf{s}}\ \mathcal{P}^*(\mathbf{s})\triangleq
\sum_{f=1}^{F}\left(p_f\mathcal{P}_f^{\text{JT}}\mathbf{1}(s_f=1)+p_f\mathcal{P}_f^{\text{PT}}\mathbf{1}(s_f\in\bar{\mathcal{N}})\right)
\end{equation}
$\centerline{\text{s.t.}\ \ (\ref{constraint-1}), (\ref{constraint-2}).}\\$

By the optimality structure of Problem 1, it can be easily verified that $\mathcal{P}^*(\mathbf{s})$ is a nondecreasing function of the cache size $C$ and SIC capability $N$. When $N=1$, we have $s_f\in\{0,1\}$ for $f\in\mathcal{F}$ and the maximization of the STP can be obtained by utilizing the MPC strategy. When $N\geq 2$, Problem 1 is a discrete optimization problem. The complexity of the exhaustive search is $O((N+1)^F)$, which is not acceptable when $N$ and $F$ become large. Therefore, we aim to provide a low complexity algorithm when $M\geq 2$.

First, we can convert the problem to a MCKP. Consider $F$ classes, each containing $N+1$ items, and a knapsack of capacity $C$. Item $n\in\mathcal{N}$ in $f$th class indicates that the cache size allocated to File $f$ is $1/q$, and item $N+1$ in the $f$th class indicates that the cache size allocated to File $f$ is 0. Each item $f\in\mathcal{N}$ in the $f$th class has a profit $a_{f,n}$ and a weight $w_{f,n}$, which correspond to the STP and the cache size of File $f$, where
\begin{equation}\label{profit-expression}
a_{f,n}\triangleq
\begin{cases}
p_f\mathcal{P}^{\text{JT}}_f, \quad\quad n=1\\
p_f\mathcal{P}^{\text{PT}}_f, \quad\quad n\in\bar{\mathcal{N}}\\\
0, \ \ \,\quad\quad\quad n=N+1
\end{cases}
\end{equation}
\begin{equation}\label{weight-expression}
w_{f,n}\triangleq
\begin{cases}
1, \ \ \quad\quad n=1\\
\frac{1}{q}, \ \ \quad\quad n\in\bar{\mathcal{N}}\\\
0, \ \,\quad\quad n=N+1
\end{cases}
\end{equation}

Note that $w_{f,n},f\in\mathcal{F}$ are the same for all $n\in\mathcal{N}$. The cache at each BS is represented by the knapsack and one item is packed from each of the $F$ classes into the knapsack. Let $x_{f,n}\in\{0,1\}$ denote whether item $q$ in the $f$th class is packed into the knapsack, where $x_{f,n}=1$ indicates that item $q$ in the $f$th class is packed into the knapsack and $x_{f,n}=0$ otherwise. Therefore, the profit sum is
\begin{equation}
\tilde{\mathcal{P}}\triangleq\sum_{f\in\mathcal{F}}\sum_{n\in\mathcal{N}}a_{f,n}x_{f,n},
\end{equation}
and the weight sum is
\begin{equation}
\sum_{f\in\mathcal{F}}\sum_{n\in\mathcal{N}}w_{f,n}x_{f,n}.
\end{equation}

Therefore, Problem 1 is converted to a MCKP, which selects one item from each class to maximize the profit sum without exceeding the cache size $C$.

\emph{Problem 2 (Equivalent Problem of Problem 1)}
\begin{equation}\label{problem-2-equation}
\mathcal{P}^*(\mathbf{x})\triangleq \max\limits_{\mathbf{x}}\ \tilde{\mathcal{P}}^*(\mathbf{x})
\end{equation}
\qquad\qquad\qquad $\text{s.t.}\ \ \sum_{f\in\mathcal{F}}\sum_{n\in\mathcal{N}}w_{f,n}x_{f,n}\leq C,\\$

\qquad\qquad\qquad\quad $\sum_{n\in\mathcal{N}}x_{f,n}=1,f\in\mathcal{F},\\$

\qquad\qquad\qquad\quad $x_{f,n}\in\{0,1\},f\in\mathcal{F},n\in\mathcal{N}.\\$

MCKP is a NP-hard problem and can be solved by utilizing two approaches, i.e., the dynamic programming and branch-bound method, with non-polynomial complexity. The drawback of these approaches lies in the dramatically increasing complexity with the increase in the number of items. Therefore, the approximate algorithm with polynomial are adopted. For instance, the performance that is no less than $1-\epsilon$ times the optimal solution can be achieved with running time polynomial to $1/\epsilon$ in the dynamic programming-based approximate algorithm, where $\epsilon\in(0,1)$. In addition, the performance that is no less than $1/2$ times the optimal solution can be achieved with complexity $O((N+1)F\log((N+1)F))$ in the greedy algorithm. In the following, we adopt a greedy algorithm to obtain a near optimal solution to Problem 2.

In order to reduce the search space of the greedy algorithm, some items could be deleted since they don't exist in an optimal solution.  Here we first introduce some definitions.

\emph{Definitions:} If two items $i$ and $j$ in the same class satisfy $w_{f,i}\leq w_{f,j}$ and $a_{f,i}\geq a_{f,j}$, then item $j$ is dominated by item $i$. If three items $i,j,k$ in the same class with $w_{f,i}<w_{f,j}<w_{f,k}$ and $a_{f,i}<a_{f,j}<a_{f,k}$ satisfy $\frac{a_{f,k}-a_{f,j}}{w_{f,k}-w_{f,j}}\geq\frac{a_{f,j}-a_{f,i}}{w_{f,j}-w_{f,i}}$, then item $j$ is LP-dominated by items $i$ and $k$.

From (\ref{profit-expression}) and (\ref{weight-expression}), we know that the indices of the undominated items are the same in each class and the item $1$ or $N+1$ is not dominated by any item in each class. Let $\mathcal{R}$ be the set of the indices of the undominated items, and denote $n^+\triangleq\min\{k|k\in\mathcal{R},k>m\}$ for all $n\in\mathcal{R}\backslash\{N+1\}$. By \cite{knapsack-problem}, we know that the dominated or LP-dominated items don't exist in an optimal solution. In other words, if item $n$ is dominated or LP-dominated by any item in the $f$th class, an optimal solution with $x_{f,n}=0$ exists. Based on these properties, we adopt the greedy algorithm to solve Problem 1. In the greedy algorithm, the prune and search algorithm is adopted to determine the set $\mathcal{R}$, then the items in $\mathcal{R}$ is sorted according to decreasing incremental efficiencies $\frac{a_{f,n}-a_{f,n^+}}{w_{f,n}-w_{f,n^+}}$. The complexity of the algorithm is $O((N+1)\log(N+1)+F(N+1)\log(F(N+1)))$, where the first term is from the reduction process and the latter term is from the sorting process. In order to reduce the complexity of the greedy algorithm, we introduce the partition algorithm, with which the complexity of the reduction process could be reduced to $O(N+1)$. Note that the partition algorithm relies on the property that finding the optimal slope, i.e., the incremental efficiency of the last item added in the greedy algorithm is sufficient to solve the MCKP. The detailed proof can be found in \cite{knapsack-problem}. The complete process is summarised below as Algorithm 1.

Note that Step 1 is conducted by utilizing the partition algorithm, as shown in Algorithm 1. In Step 2 of the partition algorithm, the pairing process is continued until all items in $N_i$ have been paired. Note that one item is unpaired if the number of items is odd.
The overall process of the MCKP is shown in Algorithm 2. In step 3, the slope $o_{f,n}$ measures the profit gain per unit weight achieved by replacing $w_{f,n}$ with $w_{f,n^+}$.  In Steps 4-8, items are selected according to their slopes in a greedy manner. In Steps 9-14, a near optimal solution to Problem 2 is constructed.

\begin{algorithm}
\begin{footnotesize}
\caption{Partition Algorithm} 
\begin{algorithmic}[1]
\For {all classes in $\mathcal{N}_f$}
\State Pair the items two by two as $(j,k)$.
\State Order each pair such that $w_{f,k}\leq w_{f,k}$ breaking ties such that $a_{f,j}\geq a_{f,k}$.
\If {$\frac{w_{f,j}}{a_{f,j}}\geq\frac{w_{f,k}}{a_{f,k}}$}
\State $\mathcal{N}_f\leftarrow N_f\backslash k$ and pair item $j$ with item $k+1$.
\EndIf
\State \textbf{end if}
\EndFor
\State $\textbf{end for}$

\For {all classes in $\mathcal{N}_f$}
\If {the class has only one item $j$}
\State $C=C-w_{f,j}$.
\EndIf
\EndFor
\State $\textbf{end for}$

\For {all pairs $(j,k)$ in $\mathcal{N}_f$}
\State $\eta_{f,j,k}=\frac{a_{f,k}-a_{f,j}}{w_{f,k}-w_{f,j}}$
\State Let $\alpha$ be the median of the slopes $\{\alpha^f_{j,k}\}$
\EndFor
\State $\textbf{end for}$

\For {$f=1,\cdot,F$}
\State $o_f(\eta)=\arg\max_{n\in \mathcal{N}_f(o^*)}\{a_{f,n}-ow_{f,n}\}$
\State $\eta_{\min,f}=\arg\min_{f\in\mathcal{N}_f(o^*)}\{w_{f,n}\}$,
\State $\eta_{\max,f}=\arg\min_{f\in\mathcal{N}_f(o^*)}\{w_{f,n}\}$
\EndFor

\If {$\sum_{f=1}^{F}w_{f,\eta_{\min,f}}\leq C<\sum_{f=1}^{F}w_{f,\eta_{\max,f}}$}
\State $o$ is the optimal slope $o^*$.
\EndIf
\If {$\sum_{f=1}^{F}w_{f,\eta_{\min,f}}>C$}
\State for all pairs $(j,k)$ in $\mathcal{N}_f$ with $o^f_{j,k}\leq o$ delete item $k$
\Else
\State for all pairs $(j,k)$ in $\mathcal{N}_f$ with $o^f_{j,k}\geq o$ delete item $j$
\EndIf
\State $\textbf{end if}$

\end{algorithmic}
\end{footnotesize}
\end{algorithm}

\begin{algorithm}
\begin{footnotesize}
\caption{Solution of Problem 1: Equation (\ref{problem-1-equation})} 
\begin{algorithmic}[1]
\State Find the set of the indices of undominated items $\mathcal{R}$ through the partition algorithm. The following indices refer to the items in $\mathcal{R}$
\State Set $x_{f,n+1}=1$, $x_{f,n}$ for all $f\in\mathcal{F}$, $n\in\mathcal{N}$, and set the weight sum $W=\sum_{f\in
\mathcal{F}}w_{f,N+1}$ and the profit sum $A=\sum_{f\in\mathcal{F}}a_{f,N+1}$.
\State For all $f\in\mathcal{F}$ and $n\in\mathcal{R}\backslash\{N+1\}$, define slope $o_{f,n}=\frac{a_{f,n}-a_{f,n^+}}{w_{f,n}-w_{f,n^+}}$. Order the slopes in $\{o_{f,n|f\in\mathcal{F},n\in\mathcal{R}\backslash\{N+1\}}\}$ in nondecreasing order. Let $o(l)$ be the $l$th largest slope.
\State $l=1$
\While {$W+w_{f,n}\leq C$}
\State Set $x_{f,n}=1$, $x_{f,n^+}=0$, and update $W=W+w_{f,n}-w_{f,n^+}$ and $A=A+a_{f,n}-a_{f,n^+}$
\State Update $l=l+1$. Let $f,n$ be the indices satisfying $o_{f,n}=o(l)$.
\EndWhile
\State $\textbf{end while}$
\If {$W=C$}
\State Set $\mathbf{x}^*=\mathbf{x}$
\Else
\State Construct a feasible solution $\bar{\mathbf{x}}\triangleq(x_{i,j})_{f\in\mathcal{F},n\in\mathcal{N}}$ to Problem 2 by setting $\bar{x}_{f,n}=1$, $\bar{x}_{i,j}=0$ for $i\in\mathcal{F},i\ne f$ or $j\in\mathcal{N},j\ne q$
\EndIf
\State $\textbf{end if}$
\State $\mathbf{x}^*=\arg\max_{\mathbf{y}\in\{\mathbf{x},\bar{\mathbf{x}}\}}\tilde{\mathcal{P}}(\mathbf{y})$
\end{algorithmic}
\end{footnotesize}
\end{algorithm}

\emph{Problem 3 (Optimization of STP in the small file size regime):}
\begin{equation}\label{problem-1-equation}
\max\limits_{\mathbf{s}}\ \mathcal{P}_{\infty}^*(\mathbf{s})\triangleq
\sum_{f=1}^{F}\left(p_f\mathcal{P}_{f,\infty}^{\text{JT}}\mathbf{1}(s_f=1)+p_f\mathcal{P}_{f,\infty}^{\text{PT}}\mathbf{1}(s_f\in\bar{\mathcal{N}})\right)
\end{equation}
$\centerline{\text{s.t.}\ \ (\ref{constraint-1}), (\ref{constraint-2}).}\\$

It is a discrete optimization problem. The indicator function hinders the solution of the optimization problem. Therefore, an equivalent problem is constructed by utilizing the optimality property of Problem 3. Since $F_c^*$ files are cached in the BSs, an auxiliary variable $F_c$ is introduced and the objective function of Problem 3 can be rewritten as
\begin{equation}\label{objective-function}
\mathcal{P}_{\infty}^{*}(\mathbf{s})\triangleq
\mathcal{P}_{f,\infty,1}(F_c)+\mathcal{P}_{f,\infty,2}^{*}(F_c),
\end{equation}
where $\mathcal{P}_{f,\infty,1}(F_c)=\sum_{f=1}^{F_c}p_f\mathcal{P}_{f,\infty}^{\text{JT}}$ and $\mathcal{P}_{f,\infty,2}(F_c)=\sum_{f=F_c+1}^{F}p_f\mathcal{P}_{f,\infty}^{\text{PT}}$.

From (\ref{objective-function}), we can observe that two types of variables are to be determined. One is the discrete variable $s_f,f\in\mathcal{F}$ and the other is the discrete variable $F_c$. Note that $F_c$ can be determined by the exhaustive search. Given $F_c$, the optimization of $s_f^*$ can be obtained by solving the following problem:

\emph{Problem 4 (Optimization of STP for given $F_c$ in the small file size regime):}
\begin{equation}\label{problem-1-equation}
\max\limits_{\mathbf{s}}\ \mathcal{P}_{\infty}^*(\mathbf{s},F_c)\triangleq
\mathcal{P}_{f,\infty,1}(F_c)+\mathcal{P}_{f,\infty,2}^{*}(F_c)
\end{equation}
\qquad\qquad\qquad $\text{s.t.}\ \ \sum_{f=F_c+1}^{F}\frac{1}{s_f}\leq C-F_c.\\$

\begin{Lemma}[Optimal solution to Problem 4]
When $\beta\rightarrow 0$, $\nu_{\text{LOS}}=1$ and $CN\leq F$, there exits $S_0>0$, such that for all $S<S_0$, $s_f^*(F_c)$ is given by
\begin{equation}\label{optimal-solution}
s_f^*(F_c)=
\begin{cases}
\frac{1}{N},&f\leq CN\\
0,&f>CN
\end{cases}
\end{equation}
and the optimal value to Problem 4 is
\begin{equation}
\begin{split}
\mathcal{P}_{\infty}^*(\mathbf{s},F_c)=&\left(1-\frac{2\left(2^{\frac{S}{s_fT}}-1\right)N!}{\alpha_{\text{LOS}}-2}
\int\limits_{0<u_1<\cdots<u_N<1}\right.\\
&\left.\frac{\text{d}u_1\cdots\text{d}u_N}{\left(1+\sum_{n=1}^{N}u_n^{-\frac{\alpha_{\text{LOS}}}{2}}\right)^{\frac{2}{\alpha_{\text{LOS}}}}}\right)\sum_{f=1}^{F_c}p_f\\
&+\left(1-\frac{S\ln2(s_f+1)}{T(2-\alpha_{\text{LOS}})}\right)\sum_{f=F_c+1}^{C-F_c}p_f.
\end{split}
\end{equation}
\end{Lemma}
\emph{Proof:} See Appendix E.

\subsection{Optimization of Average System Transmission Delay}
In this subsection, we would like to minimize the average system transmission delay by optimizing $\mathbf{s}$.

\emph{Problem 5 (Optimization of average system transmission delay):}
\begin{equation}\label{problem-3-equation}
\begin{split}
\min\limits_{\mathbf{s}}\ D^*(\mathbf{s})\triangleq
&\sum_{f=1}^{F}\left(p_f D_f^{\text{JT}}\mathbf{1}(s_f=1)+p_f D_f^{\text{PT}}\mathbf{1}(s_f\in\bar{\mathcal{N}})\right.\\
&\left.+p_f D_f^{\text{UT}}\mathbf{1}(s_f=N+1)\right)
\end{split}
\end{equation}
$\centerline{\text{s.t.}\ \ (\ref{constraint-1}), (\ref{constraint-2}).}\\$

We can convert the problem to a MCKP. Each item $f\in\mathcal{N}$ in the $f$th class has a profit $a_{f,n}$ and a weight $w_{f,n}$, which correspond to the average system transmission delay and the cache size of File $f$, where
\begin{equation}\label{profit-expression}
a_{f,n}\triangleq
\begin{cases}
p_f D^{\text{JT}}_f, \quad\quad n=1\\
p_f D^{\text{PT}}_f, \quad\quad n\in\bar{\mathcal{N}}\\\
D^{\text{UT}}_f, \quad\quad\,\,\, n=N+1
\end{cases}
\end{equation}

Therefore, Problem 5 is converted to a MCKP, which selects one item from each class to maximize the profit sum without exceeding the cache size $C$.

\emph{Problem 6 (Equivalent Problem of Problem 5)}
\begin{equation}\label{problem-2-equation}
-D^*(\mathbf{x})\triangleq \max\limits_{\mathbf{x}}\ -\tilde{D}^*(\mathbf{x})
\end{equation}
\qquad\qquad\qquad $\text{s.t.}\ \ \sum_{f\in\mathcal{F}}\sum_{n\in\mathcal{N}}w_{f,n}x_{f,n}\leq C,\\$

\qquad\qquad\qquad\quad $\sum_{n\in\mathcal{N}}x_{f,n}=1,f\in\mathcal{F}\\$

\qquad\qquad\qquad\quad $x_{f,n}\in\{0,1\},f\in\mathcal{F},n\in\mathcal{N}.\\$

By utilizing Algorithm 1 and 2, Problem 5 can be solved and the optimal value of the average system transmission delay can be obtained.

\section{Simulation Results}
In this section, we consider a mmWave cache-enabled network. We first present the the impact of the physical layer parameters on the the STP and average system transmission delay, then compare the optimal caching strategy with two baseline strategies: MPC and LDC schemes. Unless otherwise stated, the parameters are set as listed in the following table. We assume the popularity $p_f$ of the files satisfies the Zipf distribution, i.e., $p_{f}=f^{-\delta}/\sum_{f=1}^{F}f^{-\delta}$, where $\delta$ is the Zipf exponent which reflects the skewedness of the file popularity distribution. Also, the files are ranked according to their popularity: $p_1>p ... >p_{50}$.

\newcommand{\tabincell}[2]{\begin{tabular}{@{}#1@{}}#2\end{tabular}}  

\begin{table}[htbp]
\centering
\caption{\label{tab:test}System Parameters}
\begin{tabular}{|c|c|c|}
\hline
$\textbf{Parameters}$&$\textbf{Values}$\\
\hline
Transmit power & \tabincell{c}{$P=33$dBm} \\
\hline
Bias factor & $B=1$ \\
\hline
Path loss exponent & $\alpha_{\text{LOS}}=2$, $\alpha_{\text{NLOS}}=4$ \\
\hline
Density & $\lambda=50/(500^2\pi)$ \\
\hline
Blockage parameter & \tabincell{c}{$\beta=0.01$}\\
\hline
Nakagami parameter& $\nu_{\text{LOS}}=3$, $\nu_{\text{NLOS}}=2$\\
\hline
Bandwidth & \tabincell{c}{$W=1$G}\\
\hline
$\kappa_{\text{LOS}}=\kappa_{\text{NLOS}}$&$(F_c/4\pi)^2$\\
\hline
Zipf exponent & $\delta=0.6$\\
\hline
Caching capacity & $F=50$, $C=35$\\
\hline
\end{tabular}
\end{table}

\begin{figure}
  \centering
  \includegraphics[width=3.5in]{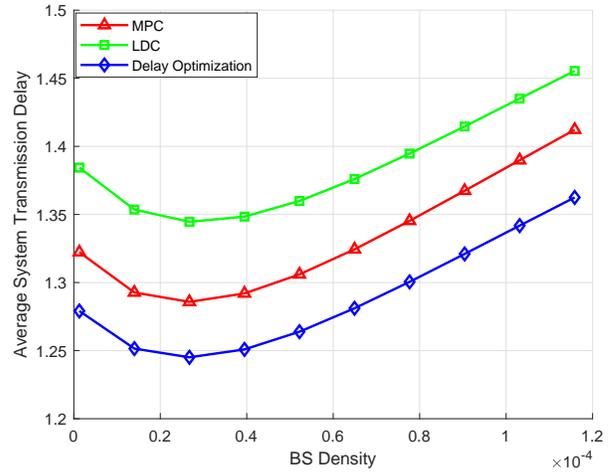}
  \caption{Average system transmission delay versus BS density $\lambda$ under different caching strategies.}\label{STP-cache-size}
\end{figure}

Fig. \ref{STP-cache-size} illustrates the effect of the cache size on the STP under different caching strategies. The numerical results match the simulation results well, thereby demonstrating the correctness of the numerical results. It can be observed that the STP increases with the cache size for three caching strategies and the proposed algorithm always outperform the MPC and LDC strategies. In addition, the gap between the three caching strategies becomes smaller when the cache size increases and the STPs become identical when the cache size is equal to 50. Note that the STP under the MPC strategy is larger than that under the LDC strategy. The reason is that the files are wholly cached in the BSs under the MPC strategy and corresponding files can be jointly transmitted by the cooperative BSs.

\begin{figure}
  \centering
  \includegraphics[width=3.5in]{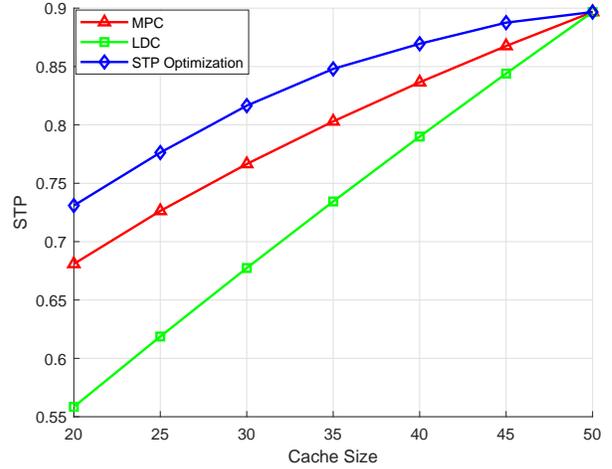}
  \caption{STP versus cache size $C$ under different caching strategies.}\label{STP-density}
\end{figure}

Fig. \ref{STP-density} illustrates the STP as functions of the density under different caching strategies. We can observe that the STPs for the three caching strategies initially increase when the density increases. However, further increase in the density causes a degradation in the STP. The reason is that when the density is relatively small, the received signal becomes stronger due to the shorten distance between $u_0$ and its serving BS. When the density increases further, the interference experienced by $u_0$ becomes larger and the enhancement in the signal power cannot compensate the increase in the interference power.

\begin{figure}
  \centering
  \includegraphics[width=3.5in]{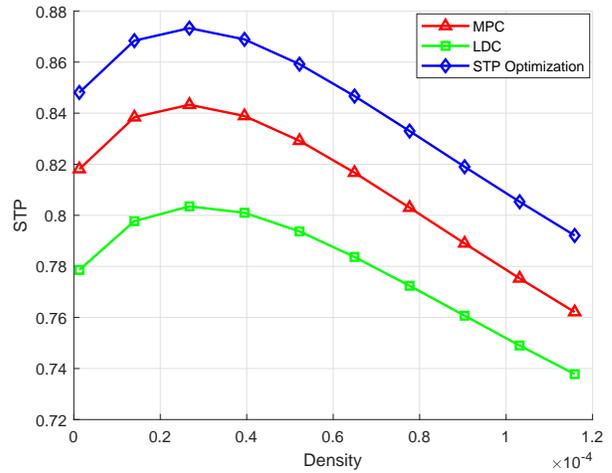}
  \caption{STP versus BS density $\lambda$ under different caching strategies.}\label{STP-Zipf-exponent}
\end{figure}

Fig. \ref{STP-Zipf-exponent} presents the effect of the Zipf exponent $\delta$ on STP. It can be seen that the STP under three caching strategies increase with the Zipf exponent $\delta$. In addition, the gain of the hybrid caching strategy over the MPC strategy approaches to zero when the Zipf exponent $\delta$ is large, indicating that only a small number of files (i.e. the most popular ones) need to be cached. In such a case, the MPC strategy can achieve the same STP as the hybrid caching strategy. In contrast, the gap between the hybrid caching strategy and the LDC strategy becomes larger when $\delta$ increases and there exists an intersection point between the two curves representing the MPC and LDC strategies. This is because when the popularity of all the files is similar or the same, the largest possible number of files should be cached. Moreover, the STP of the cache-enabled networks with MPC strategy benefits more from the BS cooperation than that with the LDC strategy.

\begin{figure}
  \centering
  \includegraphics[width=3.5in]{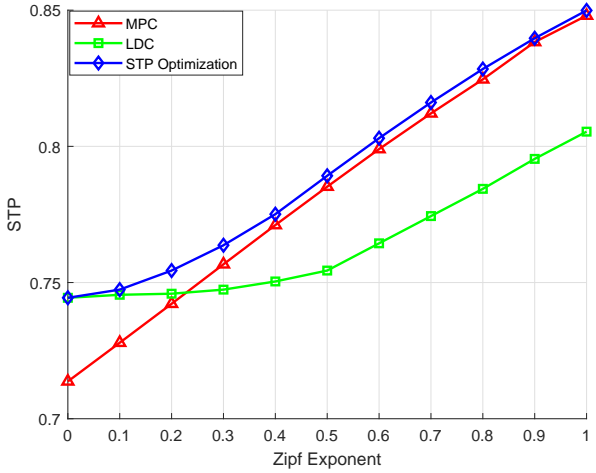}
  \caption{STP versus Zipf exponent $\delta$ under different caching strategies.}\label{ASE-cache-size}
\end{figure}

Fig. \ref{ASE-cache-size} presents the effect of the cache size on the average system transmission delay. It can be observed that the average system transmission delay under the LDC strategy is almost identical to that under the hybrid caching strategy while the average system transmission delay under the MPC strategy is larger than that under the other two strategies in the small cache size region. Note that the average system transmission delay under the MPC strategy decreases more rapidly than that under the LDC strategy and the hybrid caching strategy always performs better than the other two caching strategies. The reason is that more files can be cached in the BSs under the MPC strategy and the users are less likely to suffer from the backhaul delay.

\begin{figure}
  \centering
  \includegraphics[width=3.5in]{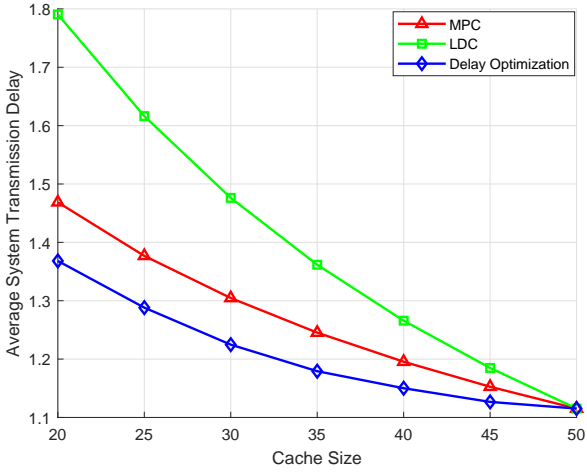}
  \caption{Average system transmission delay versus cache size $C$ under different caching strategies.}\label{ASE-density}
\end{figure}

Fig. \ref{ASE-density} illustrates the average system transmission delay as functions of the BS density $\lambda$. We can observe that the average system transmission delay for three caching strategies initially increase with the density. In addition, the STP under the MPC strategy is always larger than that under the LDC strategy.

\begin{figure}
  \centering
  \includegraphics[width=3.5in]{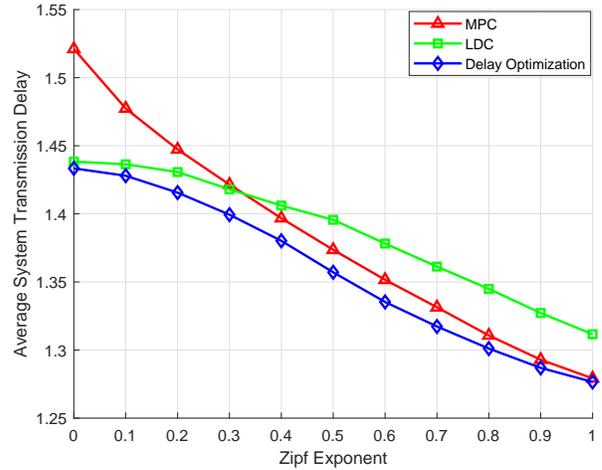}
  \caption{Average system transmission delay versus Zipf exponent $\delta$ under different caching strategies.}\label{ASE-Zipf-exponent}
\end{figure}

Fig. \ref{ASE-Zipf-exponent} illustrates the average system transmission delay as functions of the Zipf exponent $\delta$. It can be observed that the average system transmission delay under the three strategies decrease with $\delta$. The average system transmission delay under the LDC strategy is almost identical to that under the hybrid strategy and variation under the LDC strategy tends to be gentle. In contrast, the average system transmission delay under the MPC strategy is larger than that under the other two strategies in the small cache size region and decreases more rapidly than the MPC strategy. The reason is that the majority of the user requests are concentrated on fewer files when $\gamma$ increases.

\section{Conclusion}
We proposed a hybrid caching strategy for the maximization of the STP and the minimization of the average system transmission delay in the cache-enabled network. We derive STP utilizing stochastic geometry, then propose two different algorithms to obtain the optimal caching probability for maximizing the STP. Finally, the numerical results demonstrate the superiority of the hybrid caching strategy over the MPC and LDC strategies.

\section{Appendix}

\subsection{Proof of Lemma 1}
The intensity measure for the BSs can be computed as
\begin{equation}
\begin{split}
&\Lambda([0,x))
=\int_{0}^{\left(\frac{x}{\kappa_{\text{LOS}}}\right)^{\frac{1}{\alpha_\text{LOS}}}}2\pi\lambda ve^{-\beta v}\,dv\\
&+\int_{0}^{\left(\frac{x}{\kappa_{\text{NLOS}}}\right)^{\frac{1}{\alpha_\text{NLOS}}}}2\pi\lambda v\left(1-e^{-\beta v}\right)\,dv\\
&=\frac{2\pi\lambda }{\beta ^2}\left(1-e^{-\beta \left(\frac{x}{\kappa_{\text{LOS}}}\right)^{\frac{1}{\alpha_\text{LOS}}}}\left(1+
\beta \left(\frac{x}{\kappa_{\text{LOS}}}\right)^{\frac{1}{\alpha_\text{LOS}}}\right)\right)\\
&+\pi\lambda \left(\frac{x}{\kappa_{\text{NLOS}}}\right)^{\frac{2}{\alpha_\text{NLOS}}}\\
&-\frac{2\pi\lambda }{\beta ^2}\left(1-e^{-\beta \left(\frac{x}{\kappa_{\text{NLOS}}}\right)^{\frac{1}{\alpha_\text{NLOS}}}}\left(1+
\beta \left(\frac{x}{\kappa_{\text{NLOS}}}\right)^{\frac{1}{\alpha_\text{NLOS}}}\right)\right)
\end{split}
\end{equation}

The intensity for the BSs can be obtained by computing the derivative of the intensity measure as
\begin{equation}
\begin{split}
&\Lambda^{'}([0,x))=\frac{\partial\Lambda([0,x))}{\partial x}
=\frac{2\pi\lambda }{\alpha_{\text{LOS}}\kappa_{\text{LOS}}^{\frac{2}{\alpha_{\text{LOS}}}}}x^{\frac{2}{\alpha_{\text{LOS}}}-1}
e^{-\beta (\frac{x}{\kappa_{\text{LOS}}})^{\frac{1}{\alpha_{\text{LOS}}}}}\\
&+\frac{2\pi\lambda }{\alpha_{\text{NLOS}}\kappa_{\text{NLOS}}^{\frac{2}{\alpha_{\text{NLOS}}}}}x^{\frac{2}{\alpha_{\text{NLOS}}}-1}
\left(1-e^{-\beta (\frac{x}{\kappa_{\text{NLOS}}})^{\frac{1}{\alpha_{\text{NLOS}}}}}\right)
\end{split}
\end{equation}

\subsection{Proof of Theorem 1}
It is difficult to obtain the closed-form expression of a weighted sum of Nakagami random variables, we utilize the Cauchy-Schwartz's inequality to obtain the upper bound of received signal at $u_0$ as
\begin{equation}
G_0\left|\sum_{i\in\mathcal{C}}l_{i}^{-\frac{1}{2}}h_{i}\right|^2\leq G_0\sum_{i\in\mathcal{C}}l_{i}^{-1}\sum_{i\in\mathcal{C}}|h_{i}|^2
\end{equation}

Given that $u_0$ requesting File $f$ is associated with a LOS/NLOS BS, the STP can be expressed as
\begin{equation}\label{STP-pico-proof}
\begin{split}
\mathcal{P}&=\mathbbm{P}\left(\frac{PG_0\left|\sum_{i\in\mathcal{C}}l_{i}^{-\frac{1}{2}}h_{i}\right|^2}{\sigma^2+I}>T\right)\\
&\leq\mathbbm{P}\left(\frac{G_0\sum_{i\in\mathcal{C}}l_{i}^{-1}\sum_{i\in\mathcal{C}}|h_{i}|^2}{\sigma^2+I}>T\right)\\
&\overset{(a)}{=}\mathbbm{E}_{I,s}\left[\frac{\Gamma(M,s(\sigma^2+I))}{\Gamma(M)}\right]\\
&=\sum_{m=0}^{NM-1}\mathbbm{E}_{I,s}\left[e^{-s(\sigma^2+I)}\frac{(s(\sigma^2+I))^m}{m!}\right]\\
&=\mathbbm{E}_s\left[\sum_{m=0}^{NM-1}\frac{(-s)^m}{m!}\mathcal{L}^{(m)}(s)\right]
\end{split}
\end{equation}
where (a) follows from that $\sum_{i\in\mathcal{C}}|h_{i}|^2\sim\mathbb{G}\left(NM,1/M\right)$ and $s=\frac{MT}{G_0\sum_{i\in\mathcal{C}}l_{i}^{-1}}$. $\mathcal{L}(s)$ is the Laplace transform of the interference and the noise and the superscript $(m)$ denotes the $m$-th derivative of $\mathcal{L}(s)$. Due to the independence of $K$ tiers, $\mathcal{L}(s)$ can be expressed as
\begin{equation}
\begin{split}
\mathcal{L}(s)=&\exp(-s\frac{\sigma^2}{P})\mathcal{L}_{I_{\text{LOS}}}(s)\mathcal{L}_{I_{\text{NLOS}}}(s)
\end{split}
\end{equation}

We first compute the Laplace transform of the LOS interfering BSs $\mathcal{L}_{j,L}(s)$ as
\begin{equation}
\begin{split}
&\mathcal{L}_{I_{\text{LOS}}}(s)=\prod_{G}\mathbbm{E}\left[\exp\left(-s\left(\sum_{i\in\Phi\backslash \mathcal{C}}Gh_{i}L_{i}^{-1}\right)\right)\right]\\
&=\prod_{G}\prod_{i\in\Phi\backslash\mathcal{C}}\mathbbm{E}\left[\exp\left(-sGh_{i}L_{i}^{-1}\right)\right]\\
&=\prod_{G}\prod_{i\in\Phi\backslash\mathcal{C}}\frac{1}{\left(1+\frac{sG}{M}L_{i}^{-1}\right)^{M}}\\
&=\prod_{G}\exp\left(-\int_{l_{N}}^{\infty}
\left(1-\left(1+\frac{sG}{Mx}\right)^{-M}\right)\Lambda_{\text{LOS},f}(\text{d}x)\right),
\end{split}
\end{equation}

Note that $\mathcal{L}_{I_{\text{NLOS}}}(s)$ can be obtained following the similar steps. Let $\mathcal{L}(s)=\exp(\omega(s))$ and we have the 1-st order derivative of $\mathcal{L}(s)$ $\mathcal{L}^{(1)}(s)=\omega^{(1)}(s)\mathcal{L}(s)$. Therefore, $\mathcal{L}_{m}(s)$ can be derived recursively following the formula of Leibniz for the product of two functions, which is given by
\begin{equation}\label{Laplace-recursive}
\mathcal{L}^{(m)}(s)=\frac{\text{d}^{m-1}}{\text{d}s}\mathcal{L}^{(1)}(s)=\sum_{j=0}^{m-1}\binom{m-1}{j}\omega^{(m-j)}(s)\mathcal{L}^{(j)}(s).
\end{equation}

$\omega^{(j)}(s)$ can be derived as
\begin{equation}\label{nth-derivative-f}
\begin{split}
&\omega^{(j)}(s)=-\sum_{j=1}^{K}\sum_{\nu\in\{\text{LOS},\text{NLOS}\}}\sum_{G}\int_{l_{N}}^{\infty}\\
&(-1)^j(M-1)_j(Gx^{-1})^j
\left(1+\frac{sG}{Mx}\right)^{-M_{\nu}-j}\Lambda_{\nu}(\text{d}x)\\
&=-\frac{\sigma^2}{P}[2-j]^{+}-\sum_{j=1}^{K}\sum_{\nu\in\{\text{LOS},\text{NLOS}\}}\sum_G(-1)^j(M_{\nu}-1)_j\\
&Gs^{1-j}\int_{0}^{\frac{TG}{G_0l_{N}\sum_{i\in\mathcal{C}}l_{i}^{-1}}}\frac{t^{j-2}}{(1+t)^{M_{\nu}+j}}\Lambda_{\nu}(\text{d}t)\\
\end{split}
\end{equation}

Letting $x_m=\frac{1}{j!}(-s)^{j}\mathcal{L}^{(j)}(s)$, the STP can be rewritten as
\begin{equation}\label{P-xn}
\mathcal{P}=\mathbbm{E}\left[\sum_{m=0}^{NM-1}x_m\right].
\end{equation}

Substituting $x_m=\frac{1}{j!}(-s)^{j}\mathcal{L}^{(j)}(s)$ into (\ref{Laplace-recursive}), we have
\begin{equation}\label{xn-recursive}
x_m=\sum_{j=0}^{m-1}\frac{m-j}{m}\left(\frac{(-s)^{m-j}}{(m-j)!}\omega^{(m-j)}(s)\right)x_j,
\end{equation}

Letting $q_{j}\triangleq\frac{(-s)^j}{j!}\omega^{(j)}(s)$, the recursive relationship of $x_m$ in (\ref{xn-recursive}) can be rewritten as $x_m=\sum_{j=0}^{m-1}\frac{m-j}{m}q_{m-j}x_j$. In order to solve $x_m$, two power series are defined as
\begin{equation}\label{Nz-Xz}
N(z)\triangleq\sum_{m=0}^{\infty}q_{m}z^j,\ \ X(z)\triangleq\sum_{m=0}^{\infty}x_mz^m
\end{equation}

Letting $X^{(1)}(z)$ and $N^{(1)}(z)$ be the derivatives of $X(z)$ and $N(z)$, it can be easily verified that $X^{(1)}(z)=N^{(1)}(z)X(z)$. In addition, we have $X(0)=\exp(N(0))$. Therefore, the solution of (\ref{Nz-Xz}) is
\begin{equation}\label{Xz}
X(z)=\exp(N(z))
\end{equation}

Combining (\ref{P-xn}), (\ref{Nz-Xz}) and (\ref{Xz}), we have
\begin{equation}
\begin{split}
\mathcal{P}&=\mathbbm{E}\left[\sum_{m=0}^{NM-1}x_m\right]=\mathbbm{E}\left[\sum_{m=0}^{NM-1}\left.\frac{1}{m!}X^{(m)}(z)\right|_{z=0}\right]\\
&=\mathbbm{E}\left[\sum_{m=0}^{NM-1}\left.\frac{1}{m!}\frac{\text{d}^m}{\text{d}z^m}e^{N(z)}\right|_{z=0}\right]
\end{split}
\end{equation}

From \cite{norm-1-original}, the first $M$ coefficients of $\exp(N(z))$ form the first column of $\exp(\mathbf{N}_{M})$. Therefore, the STP is given by
\begin{equation}
\mathcal{P}_{\rho}=\mathbbm{E}_{\Phi}\left[\left\|\exp\left(\mathbf{N}_{M}\right)\right\|_1\right].
\end{equation}

By average over the PDF of the distance between $u_0$ and its serving BS, the STP can be obtained in (\ref{STP-pico}).

\subsection{Proof of Corollary 1}
When $\beta\rightarrow 0$, the Laplace transform of interference from the BSs can be derived as

\begin{equation}
\begin{split}
&\mathcal{L}_{I_{\infty}}(s)\\
&=\prod_{G}\mathbbm{E}\left[\exp\left(-s\left(\sum_{i\in\Phi\backslash \mathcal{C}}Gh_{i}L_{i}^{-1}\right)\right)\right]\\
&=\prod_{G}\exp\left(-\int_{l_{N}}^{R}\right.\\
&\left.\left(1-\left(1+\frac{sG}{Mx}\right)^{-M}\right)\Lambda_{\text{LOS},\infty}(\text{d}x)\right)\\
&=\pi\lambda \left(\frac{l}{\kappa_{\text{LOS}}}\right)^{\frac{2}{\alpha_{\text{LOS}}}}\\
&\left({}F_{1}\left[-\frac{2}{\alpha_{\text{LOS}}},M;1-\frac{2}{\alpha_{\text{LOS}}};\frac{TG}{MG_0\sum_{n=1}^{N}l_n^{-1}}\right]-1\right).
\end{split}
\end{equation}

Note that the intensity measure $\Lambda([0,x))$ and the intensity $\lambda([0,x))$ for $\beta\rightarrow 0$ reduces to
\begin{equation}
\Lambda_{\infty}([0,x))=\pi\lambda\left(\frac{x}{\kappa_{\text{LOS}}}\right)^{\frac{2}{\alpha_{\text{LOS}}}},
\end{equation}
\begin{equation}
\lambda_{\infty}([0,x))=\frac{2\pi\lambda}{\kappa_{\text{LOS}}^{\frac{2}{\alpha_{\text{LOS}}}}}x^{\frac{2}{\alpha_{\text{LOS}}}-1}.
\end{equation}
Thus, we have
\begin{equation}
\begin{split}
\mathcal{P}_{f,\infty}^{\text{JT}}=\idotsint\limits_{0<l_{1}<\cdots<l_{N}<\infty}\mathcal{L}_{\infty}([0,x))f_{l_{\infty}}(l)\text{d}l_{1}\cdots\text{d}l_{N},
\end{split}
\end{equation}

By using the changes of variables $u_n=\pi\lambda\left(\frac{x}{\kappa_{\text{LOS}}}\right)^{\frac{2}{\alpha_{\text{LOS}}}}$, we can get $\mathcal{P}_{f,\infty}^{\text{JT}}$ in Corollary 1.

\subsection{Proof of Theorem 2}
First, we have
\begin{equation}
\begin{split}
&\mathcal{P}_f=\mathbbm{P}\left[\bigcap\limits_{n=1,\cdots,s_f}W\left(1+\text{SINR}^{\text{PT}}_n\right)>\frac{S}{s_fT}\right]\\
&=\mathbbm{P}\left[W\left(1+\text{SINR}_1^{\text{PT}}\right)>\frac{S}{s_fT}\right]\\
&\times\prod_{n=2}^{N}\mathbbm{P}\left[W\left(1+\text{SINR}_n^{\text{PT}}\right)>\frac{S}{s_fT}\left|W\right.\right.\\
&\left.\left.\left(1+\text{SINR}_k^{\text{PT}}\right)>\frac{S}{s_fT}\right.,k=1,\cdots,n-1\right]\\
&\overset{(a)}{\approx}\prod_{n=1}^{N}\left[W\left(1+\text{SINR}_n^{\text{PT}}\right)>\frac{S}{s_fT}\right],
\end{split}
\end{equation}
where (a) is from the independence between the events $W\left(1+\text{SINR}_n^{\text{PT}}\right)>\frac{S}{s_fT},n\in{1,\cdots,N}$. Then
we derive the expression of the STP between the BS with the $n$th smallest path loss and $u_0$ after cancelling the signals from the nearer BS in $\{1,\cdots,n-1\}$. When $n$th subfile of File $f$ is transmitted by the BS, the STP $\mathcal{P}_n$ can be derived as
\begin{equation}
\begin{split}
\mathcal{P}_{f,n}&=\mathbbm{P}\left(\frac{PG_0 h_n l_{n}^{-1}}{\sigma^2+I}>T\right)\\
&=\sum_{m=0}^{M-1}\mathbbm{E}_{I,s}\left[e^{-s(\sigma^2+I)}\frac{(s(\sigma^2+I))^m}{m!}\right]\\
&=\mathbbm{E}_s\left[\sum_{m=0}^{M-1}\frac{(-s)^m}{m!}\mathcal{L}^{(m)}(s)\right]
\end{split}
\end{equation}
where $s=l_n\left(2^{\frac{S}{s_fT}}-1\right)$ and the Laplace transform of the interfering BSs can be obtained as

Since the PDF of the distance between $u_0$ and its $n$th serving BS is given by (\ref{distance-PT}), the STP can be obtained by averaging over $f_{L_n}(x)$. Combining the results from (\ref{STP-pico-proof}), the proof can be completed.

\subsection{Proof of Lemma 5}
When $s_f=0$, we have $\mathcal{P}_{f,\infty}=0$ as $S\rightarrow 0$. It remains to calculate $\mathcal{P}_{f,\infty}$ as $S\rightarrow 0$ for $s_f\in\mathcal{N}\ \{0,1\}$. When $S\rightarrow 0$, we have
\begin{equation}
{}_2F_1\left(\frac{2}{\alpha_{\text{LOS}}},1;1-\frac{2}{\alpha_{\text{LOS}}};1-2^{\frac{S}{s_fT}}\right)=1+\frac{2\left(2^{\frac{S}{s_fT}}-1\right)}{2-\alpha_{\text{LOS}}}
\end{equation}

Then we have
\begin{equation}
\begin{split}
&\mathcal{P}_{f,\infty,2}=\frac{1}{\left(1+\left(2^{\frac{S}{s_fT}}-1\right)\frac{2}{2-\alpha_{\text{LOS}}}\right)^{\frac{s_f(s_f+1)}{2}}}\\
&\overset{(a)}{=}1-\left(2^{\frac{S}{s_fT}}-1\right)\frac{2}{2-\alpha_{\text{LOS}}}\frac{s_f(s_f+1)}{2}\\
&\overset{(b)}{=}1-\frac{S\ln2(s_f+1)}{T(2-\alpha_{\text{LOS}})}
\end{split}
\end{equation}
where (a) is from $\frac{1}{(1+x)^b}=1-bx+0(x)$ as $x\rightarrow 0$.

Therefore, we have
\begin{equation}\label{small-file-size}
\begin{split}
\mathcal{P}_{f,\infty,2}(F_c)=&\sum_{f=1}^{F}p_f\mathbf{1}[s_f\ne 0,1]\\
&-\sum_{f=1}^{F}p_f\frac{S\ln2(s_f+1)}{T(2-\alpha_{\text{LOS}})}\mathbf{1}[s_f\ne 0,1].
\end{split}
\end{equation}

Substituting feasible solution $\mathbf{s}^*$ given in (\ref{optimal-solution}) into (\ref{small-file-size}), we have $\mathcal{P}_{f,\infty,2}^*(F_c)=\sum_{f=1}^{F}p_f\mathbf{1}[s_f^*\ne 0,1]-\frac{S\ln2(N+1)}{T(2-\alpha_{\text{LOS}})}\mathbf{1}[s_f^*\ne 0,1]$. On the other hand, for any feasible solution $\mathcal{s}$ to Problem 4, we have $\mathcal{P}_{f,\infty,2}(F_c)=\sum_{f=1}^{F}p_f\mathbf{1}[s_f\ne 0,1]-\frac{S\ln2(s_f+1)}{T(2-\alpha_{\text{LOS}})}\mathbf{1}[s_f\ne 0,1]$. Thus, we have
\begin{equation}\label{feasible-solution-0}
\begin{split}
&\mathcal{P}_{f,\infty,2}^*(F_c)-\mathcal{P}_{f,\infty,2}(F_c)\\
&=\left(\sum_{f=1}^{F}p_f\mathbf{1}[s_f^*\ne0,1]-\sum_{f=1}^{F}p_f\mathbf{1}[s_f\ne0,1]\right)-\frac{S\ln2(N+1)}{T(2-\alpha_{\text{LOS}})}\\
&\times\left(\sum_{f=1}^{F}p_f\mathbf{1}[s_f^*\ne0,1]-\sum_{f=1}^{F}p_f\frac{1+s_f}{1+N}\mathbf{1}[s_f\ne0,1]\right).
\end{split}
\end{equation}

For any feasible solution to Problem 3, $u_0$ can obtain at most $(C-F_c)N$ files from the $N$ cooperative BSs. Under $\mathbf{s}^*$, $u_0$ can obtain the $(C-F_c)N$ most popular files. Therefore, for any feasible solution $\mathbf{s}\ne\mathbf{s}^*$, we have
\begin{equation}\label{feasible-solution-1}
\sum_{f=1}^{F}p_f\mathbf{1}[s_f^*\ne 0,1]>\sum_{f=1}^{F}p_f\mathbf{1}[s_f\ne 0,1].
\end{equation}

If $s_f\in\bar{\mathcal{N}}$, we have $s_f\ne N$, implying $\frac{1+s_f}{1+N}\le 1$. Therefore, we have
\begin{equation}\label{feasible-solution-2}
\sum_{f=1}^{F}p_f\mathbf{1}[s_f\ne 0,1]>\sum_{f=1}^{F}p_f\frac{1+s_f}{1+N}\mathbf{1}[s_f\ne 0,1].
\end{equation}

By (\ref{feasible-solution-1}) and (\ref{feasible-solution-2}), we have
\begin{equation}
\sum_{f=1}^{F}p_f\mathbf{1}[s_f^*\ne 0,1]>\sum_{f=1}^{F}p_f\frac{1+s_f}{1+N}\mathbf{1}[s_f\ne 0,1].
\end{equation}

Based on (\ref{feasible-solution-0}), we have
\begin{equation}
\begin{split}
&\mathcal{P}_{f,\infty,2}^*(F_c)-\mathcal{P}_{f,\infty,2}(F_c)>0\\
&\Rightarrow\left(\sum_{f=1}^{F}p_f\mathbf{1}[s_f^*\ne0,1]-\sum_{f=1}^{F}p_f\frac{1+s_f}{1+N}\mathbf{1}[s_f\ne0,1]\right)\\
&\times\frac{S\ln2(N+1)}{T(2-\alpha_{\text{LOS}})}<\sum_{f=1}^{F}p_f\mathbf{1}[s_f^*\ne0,1]-\sum_{f=1}^{F}p_f\mathbf{1}[s_f\ne0,1]\\
&\Rightarrow S<\\
&\frac{\frac{T(2-\alpha_{\text{LOS}})}{S\ln2(N+1)}\left(\sum_{f=1}^{F}p_f\mathbf{1}[s_f^*\ne0,1]-\sum_{f=1}^{F}p_f\mathbf{1}[s_f\ne0,1]\right)}
{\sum_{f=1}^{F}p_f\mathbf{1}[s_f^*\ne0,1]-\sum_{f=1}^{F}p_f\frac{1+s_f}{1+N}\mathbf{1}[s_f\ne0,1]}
\end{split}
\end{equation}

\end{document}